\documentclass[journal,twoside,web]{ieeecolor}
\usepackage{generic}
\usepackage{cite}
\usepackage{amsmath,amssymb,amsfonts}
\usepackage{algorithmic}
\usepackage{graphicx}
\usepackage{textcomp}
\usepackage{multirow}
\usepackage{tabularx}
\usepackage{xcolor}
\usepackage{array,booktabs,arydshln}

\def\BibTeX{{\rm B\kern-.05em{\sc i\kern-.025em b}\kern-.08em
    T\kern-.1667em\lower.7ex\hbox{E}\kern-.125emX}}

\begin{document}
\title{Multi-objective Non-intrusive  Hearing-aid Speech Assessment Model}
\author{Hsin-Tien Chiang, Szu-Wei Fu, Hsin-Min Wang, \IEEEmembership{Senior Member, IEEE}, Yu Tsao, \IEEEmembership{Senior Member, IEEE}, and John H. L. Hansen, \IEEEmembership{Fellow, IEEE}
\thanks{This paragraph of the first footnote will contain the date on 
which you submitted your paper for review. It will also contain support 
information, including sponsor and financial support acknowledgment. For 
example, ``This work was supported in part by the U.S. Department of 
Commerce under Grant BS123456.'' }
\thanks{Hsin-Tien Chiang is with the Research Center for Information Technology Innovation, Academia Sinica, Taipei, Taiwan.}
\thanks{Szu-Wei Fu is with Nvidia, Taipei, Taiwan.}
\thanks{Hsin-Min Wang is with the Institute of Information Science, Academia Sinica, Taipei, Taiwan.}
\thanks{Yu Tsao is with the Research Center for Information Technology Innovation, Academia Sinica, Taipei, Taiwan (e-mail: yu.tsao@sinica.edu.tw).}
\thanks{John Hansen is with the Electrical and Computer Engineering, The University of Texas, Dallas, TX, USA (e-mail: john.hansen@utdallas.edu).}}

\maketitle
\begin{abstract}
\emph{\textbf{Objective:}} Without the need for a clean reference, non-intrusive speech assessment methods have caught great attention for objective evaluations. While deep learning models have been used to develop non-intrusive speech assessment methods with promising results, there is limited research on hearing-impaired subjects. This study proposes a multi-objective non-intrusive hearing-aid speech assessment model, called HASA-Net Large, which predicts speech quality and intelligibility scores based on input speech signals and specified hearing-loss patterns. \emph{\textbf{Methods:}} This paper builds on previous research on HASA-Net and improves it in several ways: (1) by considering both normal-hearing and hearing-impaired listeners; (2) by combining HASA-Net Large with other approaches such as self-supervised learning (SSL) pretraining and fine-tuning; (3) by predicting speech quality and intelligibility in a wide range of conditions, including denoising, reverberant, dereverberation, and vocoded speech, to assess its robustness; and (4) by validating the model's transferability on an out-of-domain (OOD) dataset. \emph{\textbf{Results:}} Our experiments showed the utilization of pre-trained SSL models leads to a significant boost in speech quality and intelligibility predictions compared to using spectrograms as input. Additionally, we examined three distinct fine-tuning approaches that resulted in further performance improvements. Furthermore, we demonstrated that incorporating SSL models resulted in greater transferability to OOD dataset. \emph{\textbf{Conclusion:}} The study introduces HASA-Net Large, which is a non-invasive approach for evaluating speech quality and intelligibility. HASA-Net Large utilizes raw waveforms and hearing-loss patterns to accurately predict speech quality and intelligibility levels for individuals with normal and impaired hearing and demonstrates superior prediction performance and transferability. \emph{\textbf{Significance:}} The superior prediction performance and transferability of HASA-Net Large, which highlights the model's robustness and practicality for real-world applications, represents a significant advancement in non-intrusive speech assessment for individuals with hearing impairments.
\end{abstract}


\section{Introduction}
\label{sec:introduction}
\IEEEPARstart{S}{peech} quality and intelligibility assessments serve as important tools for a variety of speech-related applications, such as speech enhancement (SE) \cite{loizou2007speech}, teleconferencing \cite{yi2022conferencingspeech}, voice conversion and text-to-speech \cite{huang2022voicemos}, and hearing aids \cite{barker20221st}. Speech quality refers to the pleasantness or naturalness of a speech signal, while speech intelligibility measures how well the content of the speech can be understood. A straightforward approach to measure speech quality or intelligibility is to conduct subjective listening tests, where a group of listeners is played speech signals and asked to score the quality or recognize the words. The mean opinion score (MOS) is a widely used criterion to assess speech quality, which ranges on a scale from one to five. Although subjective listening tests are considered the most accurate method, they are time-consuming and expensive when conducted on many subjects. Therefore, objective metrics have been proposed and used as substitutes for subjective listening tests. 
\par Objective speech assessments can be roughly divided into two categories, namely, intrusive and non-intrusive. Intrusive methods involve comparing degraded or processed speech to the clean reference to estimate perceived speech quality or intelligibility. Representative for speech quality assessment are perceptual evaluation of speech quality (PESQ) \cite{rix2001perceptual}, perceptual objective listening quality analysis (POLQA) \cite{beerends2013perceptual}, hearing aid speech quality index (HASQI) \cite{kates2014hearing1}, signal-to-distortion ratio (SDR) \cite{vincent2006performance}. For speech intelligibility assessment, some commonly used methods include articulation index (AI) \cite{french1947factors}, speech intelligibility index (SII) \cite{ansi1997methods}, speech transmission index (STI) \cite{steeneken1980physical}, short-time objective intelligibility (STOI) \cite{taal2011algorithm}, extended STOI (eSTOI) \cite{jensen2016algorithm} and hearing aid speech perception index (HASPI) \cite{kates2021hearing}. Although intrusive methods show a higher correlation with human ratings, they are not practical for real-world scenarios as clean speech may not always be available. On the other hand, non-intrusive methods calculate perceived speech quality or intelligibility directly on the degraded or processed speech without a clean reference. ITU-T P.563 \cite{malfait2006p}, speech-to-reverberation modulation ratio (SRMR) \cite{falk2010non}, and SRMR to hearing aid (SRMR-HA) \cite{suelzle2013reference} are examples of non-intrusive speech quality measures. Non-intrusive speech intelligibility measures include non-intrusive STOI (NI-STOI) \cite{andersen2017non} and modified binaural short-time objective intelligibilit (MBSTOI) \cite{andersen2018refinement}.
\par In recent years, non-intrusive models based on deep learning (DL) have shown significant progress in speech quality and intelligibility assessment. These models aim to minimize the loss between the predicted values and the ground truth values of various measures without requiring a clean reference. These approaches can be classified into two categories based on their assessment target. The first category focuses on predicting human subjective ratings. MOSNet \cite{lo2019mosnet} was designed to predict MOS for converted speech and several neural network architectures were investigated including bidirectional long short-term memory (BLSTM), convolutional neural network (CNN), and CNN-BLSTM. DNSMOS \cite{reddy2021dnsmos} used a multi-stage self-teaching approach to predict speech quality. NISQA \cite{mittag2021nisqa}, a CNN-based model, concentrated on communication network distortions and assesses speech quality across five dimensions: overall quality, noisiness, coloration, discontinuity, and loudness. MBNet \cite{leng2021mbnet} consisted of a MeanNet and BiasNet that predict the mean score of an utterance and the difference between the mean score and listener score, respectively. LDNet \cite{huang2022ldnet} directly predicted the listener score based on the input speech and listener ID, and integrates listener-dependent modeling for MOS prediction. CNN models have been employed by Andersen \emph{et al.} \cite{andersen2018nonintrusive} and Pedersen \emph{et al.} \cite{pedersen2020neural} to predict subjective intelligibility. The second category focuses on predicting objective metrics. Quality-Net \cite{fu2018quality} predicted PESQ based on BLSTM. Metric-Net \cite{yu2021metricnet} transformed the regression-based PESQ estimation to a multi-class single-label classification problem. STOI-Net \cite{zezario2020stoi} utilized CNN-BLSTM architecture with an attention mechanism to predict STOI. AMSA \cite{dong2020attention} was a unified model that predicts multiple objective speech quality and intelligibility scores, including PESQ, STOI, HASQI, and SDR.
\par Inspired by human's ability to distinguish between the quality of two speech signals regardless of their content differences, a new framework based on DL has been developed. This framework uses non-matching references to predict relative speech assessment scores, in contrast to previous DL-based reference-free methods. The NORESQA \cite{manocha2021noresqa} predicted signal-to-noise ratio (SNR) and scale-invariant signal-to-distortion ratio (Si-SNR) for quality assessment between two signals that may not be identical in terms of speech content and speaker. Additionally, NORESQA-MOS \cite{manocha2022speech} was a MOS estimation method based on the principles of NORESQA. The use of non-matching references makes these approaches applicable in real-world scenarios, as any arbitrary speech signals can be used as the reference input.
\par Self-supervised learning (SSL) has become increasingly popular in various fields, such as computer vision, natural language processing, and speech processing. These models learn feature representations from large amounts of unlabeled data and apply them to downstream tasks. In the realm of speech assessment, several studies have investigated the use of SSL. For example, Tseng \emph{et al.} \cite{tseng2021utilizing} predicted the MOS values utilizing SSL models and a listener identifier termed BiasNet to model the bias of listeners. Cooper \emph{et al.} \cite{cooper2022generalization} also used SSL models for MOS prediction, demonstrating good generalization through simple fine-tuning. Additionally, Yang \emph{et al.} \cite{yang2022fusion} proposed a a MOS prediction fusion model framework that employs seven SSL models. Furthermore, some studies have leveraged diverse acoustic information from multiple domains. For instance, Zezario \emph{et al.} used a CNN-BLSTM architecture with different acoustic features, including time-frequency domain, time-domain, and SSL embeddings as input \cite{zezario2021deep,zezario2022mbi, zezario2022mti}. Specifically, \cite{zezario2021deep} estimated PESQ, STOI, and SDI, \cite{zezario2022mbi} predicted subjective intelligibility scores for binaural hearing aid users, and \cite{zezario2022mti} estimated both subjective and objective intelligibility scores. Chen \emph{et al.} combined scattering transform and SSL models to make predictions for subjective quality and intelligibility using a Chinese dataset named TMHINT-QI \cite{chen2021inqss}.
\par Despite the recent breakthroughs of DL for speech assessment predictions, research has focused primarily on normal-hearing listeners, with limited attention given to hearing-impaired listeners.  Moreover, research focusing on hearing-impaired listeners, such as \cite{salehi2018learning,zezario2022mbi,tu2022unsupervised}, has primarily focused on predicting either quality or intelligibility instead of both criteria, and the datasets used to train these models have been limited to noisy or enhanced conditions. 
\par This study proposes HASA-Net Large, a multi-objective non-intrusive hearing-aid speech assessment model, which builds on previous work on HASA-Net \cite{chiang2021hasa}. In HASA-Net, spectrograms and hearing-loss patterns were used as input to predict speech quality and intelligibility in noisy conditions for hearing-impaired listeners. HASA-Net Large improves on this work in several ways. First, it is a general model that accounts for both normal-hearing and hearing-impaired listeners. Second, it incorporates SSL pretraining and fine-tuning approaches, which demonstrates the superiority of SSL. Third, it evaluates the model's robustness across five different speech conditions, including noisy, denoising, reverberation, dereverberation, and vocoded speech. Fourth, the model's transferability is validated using an out-of-domain (OOD) dataset in a zero-shot, few-shot, and full dataset setting.
\par The remainder of this paper is organized as follows. We present a brief introduction to HASQI and HASPI and provide an overview of prior research focused on DL-based speech assessment for hearing-impaired listeners in Section \ref{sec:relatedwork}. We present the proposed HASA-Net-2 in Section \ref{sec:hasanetlarge}. Subsequently, we conduct experiments as described in Section \ref{sec:experiment}. Finally, we conclude this work in Section \ref{sec:conclusion}.

\section{Related Work}
\label{sec:relatedwork}
\subsection{HASQI and HASPI}
In this section, we provide an introduction to the current versions of HASQI (version 2) \cite{kates2014hearing1} and HASPI (version 2) \cite{kates2021hearing}. HASQI and HASPI are objective metrics commonly used to evaluate speech quality and intelligibility for both normal and impaired hearing listeners. Both HASQI and HASPI produce scores between 0 and 1, where higher scores indicate better speech quality or intelligibility. These metrics rely on comparing the output of the model of the auditory periphery for a processed or degraded speech signal to the model output for the corresponding clean reference signal. The model of the auditory periphery \cite{kates2013auditory} used in these metrics is able to simulate normal hearing and hearing impairment that is audiogram-dependent, allowing it to represent both normal hearing and hearing-impaired conditions.
\par To compute HASQI, the reference signal used is the clean speech passed through a model of the listener's periphery. The degraded or processed signal is also passed through a peripheral model that corresponds to the degree of hearing loss. The quality assumption for HASQI is that the clean, undistorted signal processed through the listener's periphery with linear amplification \cite{byrne1986national} to compensate for any loss of audibility will result in the highest speech quality for that listener. The outputs of the auditory models are then used to measure changes in the time-frequency envelope modulation, temporal fine structure, and long-term spectrum. The nonlinear term of HASQI measures the time-frequency envelope modulation and temporal fine structure modifications, while the linear term measures the difference in the long-term spectrum. The final HASQI score is calculated as the product of the nonlinear and linear terms. More detailed information on HASQI can be found in \cite{kates2014hearing1}.
\par In HASPI, a peripheral model of the listener audiogram is used to process degraded or processed speech, while the reference signal is the clean speech that has been passed through a model of the normal auditory periphery. The assumption for HASPI is that the highest speech intelligibility will be obtained by using the sharp auditory filters and wide dynamic range that are characteristic of the normal auditory periphery \cite{kates2022overview}. The outputs of the models are analyzed using an envelope modulation-rate filterbank, and an ensemble neural network is employed to fit the subjective intelligibility scores. This is different from the original version of HASPI (version 1), which measures the outputs using a lowpass filter to determine the cepstral correlation and temporal fine structure, and employs a parametric model to fit the subjective intelligibility scores using the combined measurements of cepstral correlation and temporal fine structure. For more comprehensive details regarding both HASPI versions, please refer to \cite{kates2014hearing,kates2021hearing}. 
\par It should be noted that due to its improved performance in reverberation environments, the latest version of HASPI (version 2) has been employed to evaluate speech intelligibility in HASA-Net Large, whereas the original version of HASPI (version 1) is still utilized by HASA-Net. As for quality assessment, both HASA-Net and HASA-Net Large employ the current version of HASQI (version 2). 

\subsection{DL-based Speech Assessment Method for Hearing-aid Users}
In this section, we provide a number of DL-based methods for assessing speech assessments focusing on hearing-aid users. Given that DL-based ASR models are achieving human-level speech recognition performance and displaying comparable speech recognition patterns \cite{schadler2015matrix,fontan2017automatic,arai2020predicting}, one strategy is to utilize DL-based ASR for speech assessment. Karbasi \emph{et al.} \cite{karbasi2020non} proposed a No-Reference Intelligibility (NORI) framework based on the hidden Markov model (HMM)-based ASR model, which included two ASR-based discriminant measures for predicting speech intelligibility in noisy environments for individuals with normal and impaired hearing. 
Tu \emph{et al.} \cite{tu2022exploiting} calculated the similarity between the hidden representation from ASR models of clean reference and the processed signal to predict intelligibility in the first round Clarity Prediction Challenge \cite{graetzer2021clarity}.

\par Another approach utilizes DL models to non-intrusively estimate the output of subjective or objective metrics. For speech quality, Liang \emph{et al.} \cite{liang2023non} used a CNN to extract features from gammatone filter bank energies that were used as inputs, and employed multi-task learning to aid in predicting speech quality, along with the auxiliary task of quality classification. For speech intelligibility, MBI-Net \cite{zezario2022mbi} was developed, which includes a MSBG hearing loss model, a cross-domain feature extraction module, and a speech intelligibility prediction model to predict subjective intelligibility scores for binaural hearing-aid users. Kamo \emph{et al.} \cite{kamo2022conformer} presented a prediction model based on Conformer architecture that integrates audio, transcription, and various listener characteristics (audiogram, age, gender, etc.) to predict subjective intelligibility scores for noisy speech processed by hearing-aid users. Our previous research, HASA-Net, was able to jointly predict speech quality and intelligibility scores, specifically, HASQI and HASPI, by incorporating spectrogram and the hearing loss pattern as an extra input.

\section{HASA-Net Large}
\label{sec:hasanetlarge}
\subsection{SSL Models}
SSL has achieved great success in speech processing. SSL models learn meaningful representations from vast amounts of unlabeled data and can be categorized into three groups: generative modeling, discriminative modeling, and multi-task learning. Generative modeling relies on the network to reconstruct masked frames \cite{liu2020mockingjay,liu2021tera} or predict future frames \cite{chung2019unsupervised,chung2020improved}. Discriminative modeling employs contrastive learning \cite{schneider2019wav2vec,baevski2019vq,baevski2020wav2vec} or classify pseudo labels \cite{hsu2021hubert,chen2022wavlm} to learn meaningful speech information. Multi-task learning has been applied in \cite{ravanelli2020multi}, which learns meaningful speech information via multi-tasking objectives. The latent representations from the SSL models are used as one of the inputs to HASA-Net Large.

\subsection{Hearing-loss Patterns}
Hearing loss is detected through an audiogram, which is a graphical display showing the degrees of hearing loss at different frequency regions. The audiogram has the hearing threshold on the y-axis, measured in dB HL, and frequency on the x-axis, measured in Hz. A threshold at any frequency above 20 dB is considered as a hearing loss. We selected six features from the audiogram and used them to form a hearing-loss pattern. Each pattern describes the hearing threshold at a specific frequency. The six frequencies we chose were 250, 500, 1000, 2000, 4000, and 6000 Hz. We incorporated these hearing-loss patterns as another input of the HASA-Net Large model.

\subsection{HASA-Net Large Framework}
\label{sec:framework}
The HASA-Net Large is designed to consider both the SSL latent representations obtained from the raw waveform and the hearing-loss pattern extracted from the audiogram as inputs. These inputs are then utilized to produce the corresponding objective quality and intelligibility scores. HASA-Net Large is a modified version of HASA-Net and includes three alterations to its initial design. First, HASA-Net Large differs from HASA-Net in that it operates on the raw waveform, whereas HASA-Net utilizes time-frequency spectrograms. The second modification of HASA-Net Large involves the processing of the hearing-loss pattern. In contrast to HASA-Net, where the hearing-loss pattern is used directly, in HASA-Net Large, it undergoes additional processing. Specifically, the hearing-loss pattern is passed through a dense layer, which increases its dimensionality from 6 to 256. The third modification in HASA-Net Large involves a different approach to combining the inputs before feeding them into the BLSTM layer. In HASA-Net, the input to the BLSTM layer was obtained by concatenating the time-frequency spectrograms and hearing-loss pattern. In contrast, in HASA-Net Large, the latent representations obtained from the SSL model and the hearing-loss pattern processed through the dense layer are combined using addition, resulting in merged features that are passed into the BLSTM layer. 

\begin{figure}[t!]
\begin{minipage}[b]{0.9\linewidth}
  \centering
  \centerline{\includegraphics[width=8cm]{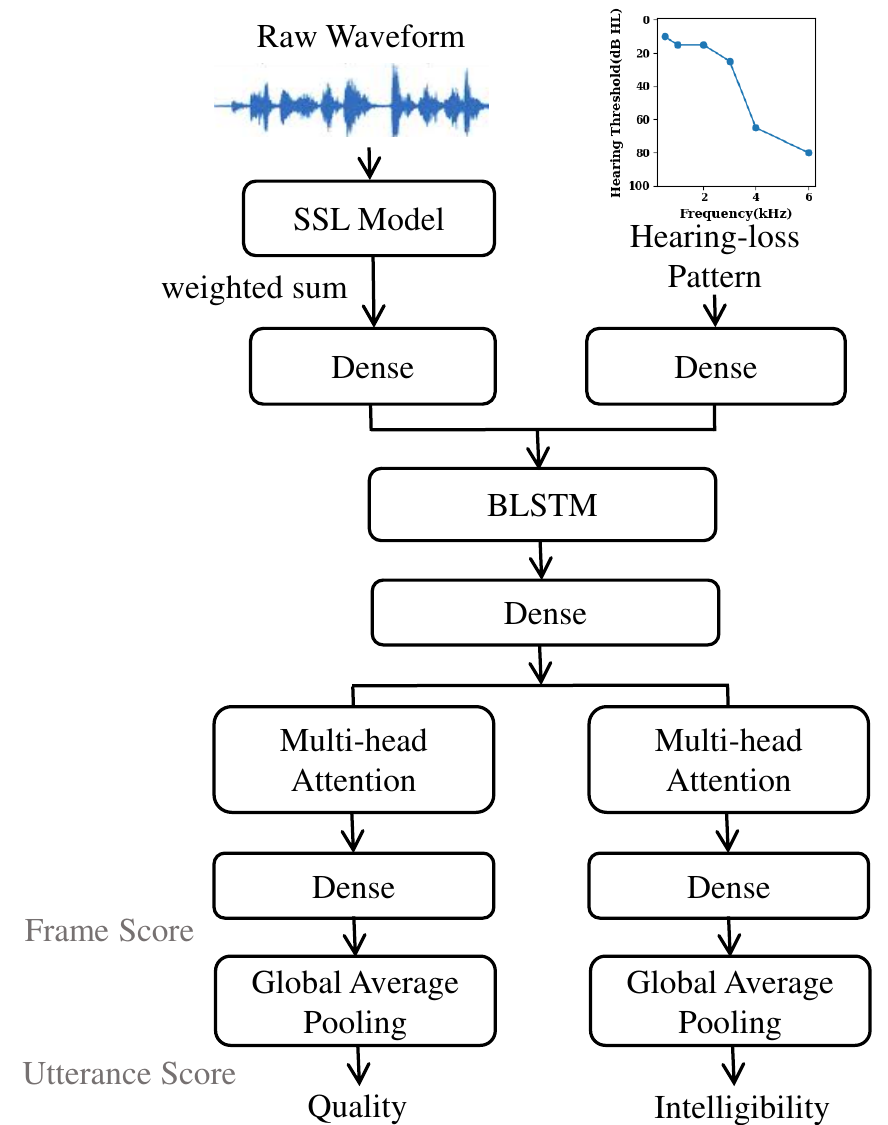}}
\end{minipage}
\caption{Architecture of the HASA-Net Large.}
\label{fig:hasanet2}
\end{figure}

\par The overall framework of HASA-Net Large is depicted in Fig. \ref{fig:hasanet2}. The left branch of the input corresponds to the SSL representation, which is obtained by feeding the raw waveform into an SSL model. The SSL representation is a weighted sum of the representations from all transformer encoder layers, and it is passed through a dense layer to have a dimensionality of 256. The right branch of the input represents the hearing-loss pattern, which is obtained from the audiograms and describes the hearing threshold at six specific frequencies. This pattern is passed through a fully connected layer, increasing its dimensionality from 6 to 256. After merging the SSL representation and the hearing-loss feature, the resulting merged features are fed into a stack of one bidirectional LSTM with 100 nodes, followed by one dense layer with 128 ReLU nodes.
Finally, the output of the dense layer is split into two separate tasks, quality estimation and intelligibility prediction. For each task, a multi-head attention mechanism, a dense layer consisting of one node with a sigmoid function, and a global average pooling are applied to generate the final prediction results. The outputs of the dense layer and the global average pooling are frame-level prediction and utterance-level prediction, respectively.
\par The total loss is the sum of the losses for the two tasks, and in each task, the loss function for each utterance is the summation of utterance-level loss and averaged frame-wise loss. The objective function is denoted as follows:
\begin{displaymath}\label{eq:totalloss}
\begin{aligned}
L_{Total} = L_{Quality}+L_{Intelligibility}
\end{aligned}
\end{displaymath}
\begin{displaymath}\label{eq:qualityloss}
\begin{aligned}
L_{Quality} = \frac{1}{N}\sum_{n=1}^{N}[(\hat{Q_n}-Q_n)^2 + \frac{1}{T_n} \sum_{t=1}^{T_n}(\hat{Q_n}-q_{n,t})^2 ]
 \end{aligned}
\end{displaymath}
\begin{displaymath}\label{eq:intelligibilityloss}
\begin{aligned}
L_{Intelligibility} = \frac{1}{N}\sum_{n=1}^{N}[(\hat{I_n}-I_n)^2 + \frac{1}{T_n} \sum_{t=1}^{T_n}(\hat{I_n}-i_{n,t})^2 ]
\end{aligned}
\end{displaymath}
where \{$\hat{Q_n}$, $Q_n$\} and \{$\hat{I_n}$, $I_n$\} represent the true and estimated scores for the $n$-th utterance of quality and intelligibility, respectively, while $N$ represents the total training utterances and $T_n$ is the number of frames in utterance $n$. Meanwhile, $q_{n,t}$ and $i_{n,t}$ denote the estimated quality and intelligibility frame score of the $t$-th frame of utterance $n$. 

\section{Experiments}
\label{sec:experiment}
\subsection{Dataset Creation}
We utilized the clean speech from the VCTK-DEMAND corpus \cite{valentini2016investigating} as in-domain data and the TIMIT corpus \cite{garofolo1988getting} as OOD data. Both in- and out-of-domain datasets used in the experiments comprised a variety of speech types, including noisy, enhanced, reverberation, dereverberation, and vocoded speech. We provide a comprehensive description of the setup and data creation process for each dataset below.
\subsubsection{In-domain dataset}
Initially, the training set included 11,572 utterances from 28 distinct speakers. We started by dividing the VCTK-DEMAND training set into two parts. We selected two speakers, p226 and p287, to form a validation set of 770 utterances, while the remaining utterances (10,802 utterances) were utilized to train DL-based enhancement or dereverberation models, which we called training set 1. The testing set comprised 824 spoken utterances from two distinct speakers. It is noteworthy that the utterances that were utilized to train the DL-based enhancement and dereverberation models were not considered in the HASA-Net Large experiments. On the other hand, we combined the validation and testing sets to establish a new training set comprising 1,594 utterances, which we called training set 2, for use in the HASA-Net Large experiments.  
\par The VCTK-DEMAND corpus was the source of the noisy speech, and it included four signal-to-noise ratios (SNRs) in the training set (15, 10, 5, and 0 dB) and four SNRs in the testing set (17.5, 12.5, 7.5, and 2.5 dB). We trained MetricGAN+ \cite{fu2021metricgan+} using the data in training set 1 to generate enhanced speech. The creation of the reverberation and dereverberation sets followed the setting outlined in \cite{fu2022metricgan}, using the $AddReverb$ function in the SpeechBrain \cite{ravanelli2021speechbrain} toolkit. Specifically, we generated reverberation speech by convolving clean speech from the training set 1 and validation set with 315 room impulse response (RIR) data, and clean speech from the testing set with 10 RIRs. Each clean utterance was convolved with only one RIR and a corresponding $rir\underline{~}scale \underline{~}factor$. Additional information about this process can be found in \cite{fu2022metricgan}. The dereverberation speech was produced using MetricGAN-U, which was trained using the reverberation data from training set 1. To create the vocoded sets, we applied the tone vocoder and noise vocoder on the training set 2. Specifically, we generated half of the data using the tone vocoder, while the other half were produced using the noise vocoder.
\par The in-domain dataset used in the HASA-Net Large consists of training set 2, which includes the validation and testing sets that were not used for training DL-based enhancement or dereverberation models. Each utterance in the VCTK-DEMAND dataset is associated with five different conditions, including noisy, reverberation, enhancement, dereverberation, and vocoded. In total, the in-domain dataset contains 7,970 (1,594 $\times$ 5) utterances.
\subsubsection{OOD dataset}
The training set comprised 4,680 utterances from 90 speakers, and the testing set comprised 1,560 utterances from 30 speakers. We selected 200 clean utterances from the TIMIT testing set to train DL-based enhancement or dereverberation models. It is important to note that the 200 utterances mentioned above were not used in the HASA-Net Large experiments.
\par To produce the noisy and enhanced sets, we used the identical noise signal from the VCTK-DEMAND dataset to each clean utterance, at the same SNR levels as used for the VCTK-DEMAND dataset, resulting in the generation of noisy speech. The BLSTM-based SE model was trained on 4,000 utterances generated by randomly corrupting the 200 clean utterances with 5 different noises from the DEMAND database at 4 SNRs (2.5, 7.5, 12.5, and 17.5 dB). The enhanced speech was then generated using this SE model. Reverberation speech was created by convolving 4,680 clean speech utterances from the training set with 315 RIRs, and 1,360 clean speech utterances from the testing set with 10 RIRs, which were the same as those used in the in-domain dataset setup. To obtain dereverberation data, a BLSTM model was trained using 5,000 utterances. These trainind data was created by applying convolution to the 200 utterances with 5 RIRs at 5 distinct $rir\underline{~}scale \underline{~}factor$ (0.75, 0.85, 0.95, 1.05, and 1.15). To create the vocoded version of the data, both the tone vocoder and noise vocoder were employed. Half of the data was generated using the tone vocoder, while the remaining half was produced using the noise vocoder.
\par After removing the 200 utterances used for training DL-based enhancement or dereverberation models from the testing set of TIMIT corpus, there remained 4,620 utterances from the training set and 1,260 utterances from the testing set of the TIMIT corpus. Each utterance corresponded to five different conditions (noisy, reverberation, enhancement, dereverberation, or vocoded). Therefore, the total number of utterances in the OOD training and testing sets was 23,100 (4,620 $\times$ 5) and 6,300 (1,260 $\times$ 5), respectively.

\subsection{Audiograms}
\begin{figure}[t!]
  \centering
  \centerline{\includegraphics[width=8cm]{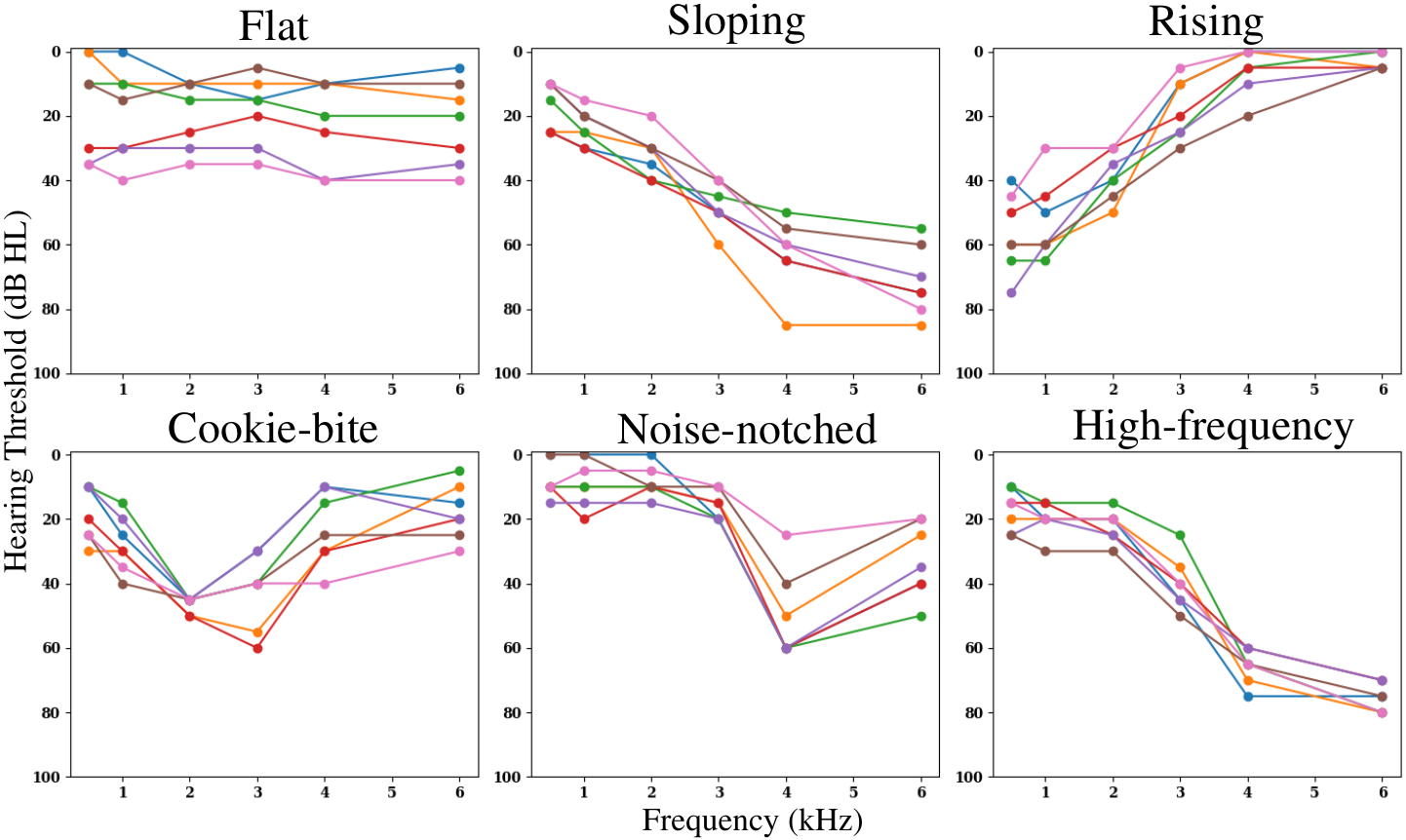}}
\caption{The hearing loss audiograms for the evaluation of speech quality and intelligibility.}
\label{fig:audiogram}
\end{figure}

We employed the identical audiograms for hearing loss that were utilized in \cite{chiang2021hasa}These audiograms are divided into six categories: flat, sloping, rising, cookie-bite, noise-notched, and high-frequency, and are depicted in Figure \ref{fig:audiogram}. Each category comprises seven different audiograms. Furthermore, we also took into account a scenario where all frequencies had a 0 dB HL, indicating normal hearing. As a result, there were a total of 43 hearing-loss audiograms, which included 7 audiograms in each of the 6 hearing loss categories, plus 1 audiogram for normal hearing. 

\par In the HASA-Net Large experiments, the audiograms were split into different sets for training, validation, and testing. The training set contained 30 patterns, with 5 audiograms in each category, while the test set had 12 patterns, with 2 audiograms in each category. Additionally, 12 patterns were selected from the training set to create a validation set, with 2 audiograms in each category. The audiogram with 0 dB HL, representing normal hearing, was used in all three sets. The audiograms in the training and test sets did not overlap, except for normal hearing, while the training and validation sets shared the same audiograms. Overall, the training, validation, and test sets comprised 31, 13, and 13 audiograms, respectively.

\subsection{Results}
\subsubsection{Evaluation of pre-trained SSL models}

\begin{table*}[t]
    \caption{Comparison of Quality and Intelligibility Prediction Results between HASA-Net and HASA-Net Large with Pre-trained WavLM Model.}
    \centering
    \begin{tabular}{lcccccc}
    \toprule[0.4mm]    
    &\multicolumn{3}{c}{Quality}&\multicolumn{3}{c}{Intelligibility}\\
    \cline{2-4}\cline{5-7}
    Model & MSE$\downarrow$ & LCC$\uparrow$ & SRCC$\uparrow$ & MSE$\downarrow$ & LCC$\uparrow$ & SRCC$\uparrow$\\
    \hline
    HASA-Net & 0.007 & 0.947& 0.958 & 0.026 & 0.737 & 0.756 \\
    \hdashline
    HASA-Net Large & \multirow{2}{*}{0.006} & \multirow{2}{*}{0.951}& \multirow{2}{*}{0.960} & \multirow{2}{*}{0.024} & \multirow{2}{*}{0.733} & \multirow{2}{*}{0.764} \\
    (WavLM-LL)  \\
    HASA-Net Large  & \multirow{2}{*}{\textbf{0.005}} & \multirow{2}{*}{\textbf{0.955}} & \multirow{2}{*}{\textbf{0.969}} & \multirow{2}{*}{\textbf{0.019}} & \multirow{2}{*}{\textbf{0.808}} & \multirow{2}{*}{\textbf{0.829}} \\
    (WavLM-WS) \\
    \bottomrule[0.4mm]
    \label{table:pretrain}
    \end{tabular}%
\end{table*}

\begin{table*}[t]
    \caption{Detailed Evaluation of Quality Prediction between HASA-Net and HASA-Net Large under Varied Conditions.}
    \centering
    \begin{tabular}{lccccccccc}
    \toprule[0.4mm]    
    & \multicolumn{3}{c}{\multirow{2}{*}{HASA-Net}}&\multicolumn{3}{c}{HASA-Net Large}&\multicolumn{3}{c}{HASA-Net Large}\\
    & \multicolumn{3}{c}{}& \multicolumn{3}{c}{(WavLM-LL)} & \multicolumn{3}{l}{(WavLM-WS)} \\
    \cline{2-4}\cline{5-7}\cline{8-10}
    condition & MSE$\downarrow$ & LCC$\uparrow$ & SRCC$\uparrow$ & MSE$\downarrow$ & LCC$\uparrow$ & SRCC$\uparrow$ 
    & MSE$\downarrow$ & LCC$\uparrow$ & SRCC$\uparrow$\\
    \hline
    noisy & 0.008 & 0.871 & 0.936 & 0.008 & 0.836 & 0.937 & 0.005 & 0.878 & 0.958 \\
    enhanced & 0.010 & 0.888 & 0.921 & 0.006 & 0.935 & 0.952 & 0.004 & 0.955 & 0.967 \\
    reverberation & 0.009 & 0.843 & 0.881 & 0.006 & 0.883 & 0.921 & 0.004 & 0.909 & 0.941 \\
    dereverberation & 0.007 & 0.852 & 0.885 & 0.006 & 0.861 & 0.900 & 0.005 & 0.890 & 0.922 \\
    vocoded & 0.004 & 0.908 & 0.967 & 0.005 & 0.912 & 0.957 & 0.004 & 0.933 & 0.970 \\
    \bottomrule[0.4mm]
    \end{tabular}
    \label{table:detailedQ}
\end{table*}

\begin{table*}[t]
    \caption{Detailed Evaluation of Intelligibility Prediction between HASA-Net and HASA-Net Large under Varied Conditions.}
    \centering
    \begin{tabular}{lccccccccc}
    \toprule[0.4mm]    
    & \multicolumn{3}{c}{\multirow{2}{*}{HASA-Net}}&\multicolumn{3}{c}{HASA-Net Large}&\multicolumn{3}{c}{HASA-Net Large}\\
    & \multicolumn{3}{c}{}& \multicolumn{3}{c}{(WavLM-LL)} & \multicolumn{3}{c}{(WavLM-WS)} \\
    \cline{2-4}\cline{5-7}\cline{8-10}
    condition & MSE$\downarrow$ & LCC$\uparrow$ & SRCC$\uparrow$ & MSE$\downarrow$ & LCC$\uparrow$ & SRCC$\uparrow$ & MSE$\downarrow$ & LCC$\uparrow$ & SRCC$\uparrow$ \\
    \hline
    noisy & 0.025 & 0.653 & 0.678 & 0.022 & 0.653 & 0.707 & 0.015 & 0.740 & 0.811 \\
    enhanced & 0.021 & 0.641 & 0.642 & 0.018 & 0.664 & 0.701 & 0.014 & 0.738 & 0.778 \\
    reverberation & 0.022 & 0.749 & 0.763 & 0.021 & 0.759 & 0.782 & 0.014 & 0.837 & 0.855 \\
    dereverberation & 0.028 & 0.801 & 0.792 & 0.028 & 0.787 & 0.794 & 0.022 & 0.841 & 0.841 \\
    vocoded & 0.028 & 0.665 & 0.726 & 0.028 & 0.673 & 0.722 & 0.024 & 0.711 & 0.773 \\
    \bottomrule[0.4mm]
    \end{tabular}
    \label{table:detailedI}
\end{table*}

As described in section~\ref{sec:framework}, the HASA-Net Large model takes both the SSL latent representations extracted from the raw waveform and the hearing-loss pattern obtained from the audiogram as inputs, and produces the corresponding objective quality and intelligibility scores. The ground-truth values for quality and intelligibility were determined using HASQI and HASPI, respectively. All the utterances were presented at 65 dB sound pressure level (SPL) for listeners with normal hearing. For hearing-impaired listeners, the stimuli were amplified using the National Acoustics Laboratories revised (NAL-R) \cite{byrne1986national} linear fitting prescriptive formula, which was based on their individual audiograms. The evaluation criteria included mean square error (MSE), linear correlation coefficient (LCC), and Spearman's rank correlation coefficient (SRCC). 

\par To evaluate the performance of HASA-Net Large on the in-domain dataset, a $k$-fold cross-validation with $k=5$ was employed due to the limited number of unique clean speech utterances (1594 utterances). In $k$-fold cross-validation, the training set data (consisting of 1594 $\times$ 5 utterances) was randomly divided into $k$ partitions, and one of the partitions was kept as the test set while the remaining $k-1$ partitions were used for training. Each utterance was matched with three audiograms: 2 out of 13 selected from the training set and the normal-hearing audiogram. This pairing resulted in a total of 1594 $\times$ 4 $\times$ 3 combinations for training. During testing, each utterance was paired with three audiograms: 2 out of 12 selected from the testing set and the normal-hearing audiogram. This pairing resulted in a total of 1594 $\times$ 3 combinations. The cross-validation process was repeated $k$ times, and the outcomes were averaged to obtain an overall performance estimation. Each of the $k$ partitions was used once as the test set, and the $k$ results were then averaged to obtain the overall performance estimation. 
\par In our evaluation, we compare the performance of two models: HASA-Net Large, which integrates a pre-trained SSL model, and HASA-Net, which uses spectrograms as input and serves as our baseline. We use the pre-trained WavLM Large \cite{chen2022wavlm} as the SSL model and investigate two methods for integrating its representation into downstream tasks: using the representation from the last layer (LL) or computing a weighted sum of the representations from all transformer encoder layers with learnable weights (WS). As a result, we have two versions of HASA-Net Large: HASA-Net Large (WavLM-LL) and HASA-Net Large (WavLM-WS). For training both versions of HASA-Net Large, we utilize the Adam optimizer with a learning rate set to $10^{-4}$.

\par Table \ref{table:pretrain} shows that both versions of HASA-Net Large outperform the baseline HASA-Net, indicating the superiority of SSL representations. Moreover, comparing the two versions of HASA-Net Large, we find that HASA-Net (WavLM-WS) achieves higher correlation values and lower MSE values than HASA-Net (WavLM-LL). This suggests that each transformer layer in the SSL models contains valuable information, and fully leveraging the information from different layers is crucial for optimal performance.

\begin{table*}[t]
    \caption{Comparing Three Fine-tuning Approaches (PF, EF, and 2-stage FT) for Quality and Intelligibility Prediction.}
    \centering
    \begin{tabular}{lcccccc}
    \toprule[0.4mm]    
    HASA-Net Large&\multicolumn{3}{c}{Quality}&\multicolumn{3}{c}{Intelligibility}\\
    \cline{2-4}\cline{5-7}
    (WavLM-WS) & MSE$\downarrow$ & LCC$\uparrow$ & SRCC$\uparrow$ & MSE$\downarrow$ & LCC$\uparrow$ & SRCC$\uparrow$\\
    \hline
    pre-trained & 0.005 & 0.969 & 0.955 & 0.019 & 0.829 & 0.808 \\
    \hdashline
    PF & 0.003 & 0.979 & 0.974 & 0.019 & 0.823 & 0.804 \\
    EF & 0.003 & 0.980 & 0.970 & 0.018 & 0.848 & 0.841 \\
    2-stage FT & \textbf{0.002} & \textbf{0.987} & \textbf{0.983} & \textbf{0.013} & \textbf{0.885}  & \textbf{0.869} \\
    \bottomrule[0.4mm]
    \end{tabular}
    \label{table:finetuning}
\end{table*}

\begin{table*}[t]
    \caption{Comparison of Quality and Intelligibility Prediction Results in Zero-shot, Few-shot, and Full-shot Settings with Varying Training Data Sizes.}
    \centering
    \begin{tabular}{llcccccc}
    \toprule[0.4mm]    
    &&\multicolumn{3}{c}{Quality}&\multicolumn{3}{c}{Intelligibility}\\
    \cline{3-5}\cline{6-8}
    training data size & Model & MSE$\downarrow$ & LCC$\uparrow$ & SRCC$\uparrow$ & MSE$\downarrow$ & LCC$\uparrow$ & SRCC$\uparrow$ \\
    \hline
    \multirow{3}{*}{zero-shot} & HASA-Net & 0.111 & 0.439& 0.473 & 0.023 & 0.240& 0.458 \\
    & HASA-Net Large & \multirow{2}{*}{\textbf{0.028}} & \multirow{2}{*}{\textbf{0.765}} & \multirow{2}{*}{\textbf{0.775}} & \multirow{2}{*}{\textbf{0.013}} & \multirow{2}{*}{\textbf{0.615}} & \multirow{2}{*}{\textbf{0.802}} \\
    & (WavLM-WS)  \\
    \hdashline
    \multirow{3}{*}{100} & HASA-Net & 0.035 & 0.748 & 0.703 & 0.016 & 0.416 & 0.544 \\
    & HASA-Net Large & \multirow{2}{*}{\textbf{0.007}} & \multirow{2}{*}{\textbf{0.951}} & \multirow{2}{*}{\textbf{0.870}} & \multirow{2}{*}{\textbf{0.010}} & \multirow{2}{*}{\textbf{0.652}} & \multirow{2}{*}{\textbf{0.817}} \\
    & (WavLM-WS)  \\
    \hdashline
    \multirow{3}{*}{400} & HASA-Net & 0.033 & 0.779 & 0.740 & 0.015 & 0.479 & 0.636 \\
    & HASA-Net Large & \multirow{2}{*}{\textbf{0.004}} & \multirow{2}{*}{\textbf{0.972}} & \multirow{2}{*}{\textbf{0.933}} & \multirow{2}{*}{\textbf{0.010}} & \multirow{2}{*}{\textbf{0.669}} & \multirow{2}{*}{\textbf{0.833}} \\
    & (WavLM-WS)  \\
    \hdashline
    \multirow{3}{*}{1,600} & HASA-Net & 0.021 & 0.849 & 0.790 & 0.014 & 0.516 & 0.677 \\
    & HASA-Net Large & \multirow{2}{*}{\textbf{0.003}} & \multirow{2}{*}{\textbf{0.976}} & \multirow{2}{*}{\textbf{0.944}} & \multirow{2}{*}{\textbf{0.010}} & \multirow{2}{*}{\textbf{0.690}} & \multirow{2}{*}{\textbf{0.813}} \\
    & (WavLM-WS)  \\
    \hdashline
    \multirow{3}{*}{6,400} & HASA-Net &0.014 & 0.903 & 0.848 & 0.013 & 0.588 & 0.740  \\
    & HASA-Net Large & \multirow{2}{*}{\textbf{0.003}} & \multirow{2}{*}{\textbf{0.981}} & \multirow{2}{*}{\textbf{0.967}} & \multirow{2}{*}{\textbf{0.009}} & \multirow{2}{*}{\textbf{0.707}} & \multirow{2}{*}{\textbf{0.833}} \\
    & (WavLM-WS)  \\
    \hdashline
    \multirow{3}{*}{full-shot} & HASA-Net & 0.011& 0.923 & 0.867 & 0.010 & 0.653 & 0.769 \\
    & HASA-Net Large & \multirow{2}{*}{\textbf{0.002}} & \multirow{2}{*}{\textbf{0.986}} & \multirow{2}{*}{\textbf{0.980}} & \multirow{2}{*}{\textbf{0.008}} & \multirow{2}{*}{\textbf{0.756}} & \multirow{2}{*}{\textbf{0.866}} \\
    & (WavLM-WS)  \\
    \bottomrule[0.4mm]
    \label{table:trans}
    \end{tabular}%
\end{table*}

\begin{table*}[t]
    \caption{Performance of quality prediction of different types of hearing loss with varying training data sizes.}
    \centering
    \begin{tabular}{lcccccccccccc}
    \toprule[0.4mm]    
    training data size & \multicolumn{3}{c}{zero-shot}&\multicolumn{3}{c}{100}&\multicolumn{3}{c}{1,600}&\multicolumn{3}{c}{full-shot}\\
    \cline{1-13}
     & MSE$\downarrow$ & LCC$\uparrow$ & SRCC$\uparrow$ & MSE$\downarrow$ & LCC$\uparrow$ & SRCC$\uparrow$ 
    & MSE$\downarrow$ & LCC$\uparrow$ & SRCC$\uparrow$ 
    & MSE$\downarrow$ & LCC$\uparrow$ & SRCC$\uparrow$ \\
    \hline
    flat & 0.036 & 0.736 & 0.747 & 0.006 & 0.963 &0.863 & 0.003 & 0.982 &0.940 & 0.002 & 0.989 & 0.978\\
    sloping & 0.028 & 0.787 & 0.793 & 0.008 & 0.959 & 0.902 & 0.003 & 0.979 &0.950 &0.002 &0.988 &0.986\\
    rising & 0.027 & 0.769 & 0.771 & 0.010 & 0.932 & 0.859 & 0.004 &0.967 &0.923 &0.003 &0.980 &0.969\\
    cookie-bite & 0.030 & 0.761 & 0.740 &0.006 &0.957 &0.865 & 0.003 &0.978 &0.912 &0.002 &0.986 &0.975 \\
    noise-notched & 0.020 & 0.846 & 0.861 & 0.005 & 0.955 & 0.871 & 0.004 &0.975 &0.903 & 0.001 &0.989 & 0.968\\
    high-frequency & 0.046 & 0.668 & 0.726 & 0.006 & 0.973 &0.871 & 0.003 & 0.983 &0.940 & 0.002 &0.991 &0.981\\
    normal & 0.014 & 0.938 & 0.882 & 0.008 &0.938 &0.887 & 0.004 &0.969 &0.937 & 0.001 &0.985 &0.973\\
    \bottomrule[0.4mm]
    \end{tabular}
    \label{table:detailedQ_diffsize}
\end{table*}

\begin{table*}[t]
    \caption{Performance of intelligibility prediction of different types of hearing loss with varying training data sizes.}
    \centering
    \begin{tabular}{lcccccccccccc}
    \toprule[0.4mm]    
    training data size & \multicolumn{3}{c}{zero-shot}&\multicolumn{3}{c}{100}&\multicolumn{3}{c}{1,600}&\multicolumn{3}{c}{full-shot}\\
    \cline{1-13}
     & MSE$\downarrow$ & LCC$\uparrow$ & SRCC$\uparrow$ & MSE$\downarrow$ & LCC$\uparrow$ & SRCC$\uparrow$ 
    & MSE$\downarrow$ & LCC$\uparrow$ & SRCC$\uparrow$ 
    & MSE$\downarrow$ & LCC$\uparrow$ & SRCC$\uparrow$ \\
    \hline
    flat & 0.017 & 0.644 & 0.819 & 0.012 & 0.636 &0.838 & 0.012 & 0.657 &0.826 & 0.009 & 0.733 & 0.855\\
    sloping & 0.011 & 0.562 & 0.472 & 0.009 & 0.580 & 0.510 & 0.010 & 0.640 &0.526 &0.008 &0.681 &0.616\\
    rising & 0.023 & 0.560 & 0.737 & 0.019 & 0.655 & 0.795 & 0.016 &0.707 &0.759 &0.012 &0.791 &0.818\\
    cookie-bite & 0.014 & 0.457 & 0.565 &0.010 &0.521 &0.583 & 0.010 &0.562 &0.546 &0.008 &0.627 &0.656 \\
    noise-notched & 0.004 & 0.333 & 0.405 & 0.005 & 0.160 & 0.460 & 0.005 &0.315 &0.449 & 0.004 &0.312 & 0.490\\
    high-frequency & 0.020 & 0.585 & 0.477 & 0.014 & 0.651 & 0.544 & 0.014 & 0.693 &0.564 & 0.011 &0.756 &0.662\\
    normal & 0.005 & 0.534 & 0.614 & 0.0002 &0.641 &0.616 & 0.0003 & 0.743 & 0.594 & 0.0002 & 0.780 & 0.609\\
    \bottomrule[0.4mm]
    \end{tabular}
    \label{table:detailedI_diffsize}
\end{table*}

Table \ref{table:detailedQ_diffsize} and \ref{table:detailedI_diffsize} present additional information regarding the performance of quality and intelligibility prediction, specifically focusing on various types of hearing loss. Once again, we observe a consistent trend where the correlation values show an upward trajectory, while the MSE decreases as the training data size increases for all types of hearing loss. In terms of quality prediction, each hearing loss type exhibits a notable improvement when there are 100 training data samples available. This suggests that a relatively small but sufficient training dataset size can lead to a significant enhancement in the performance of the quality prediction model for different types of hearing loss. In contrast, the impact of increasing the training data size is not as substantial for intelligibility prediction compared to quality prediction.

\par We also examine the detailed evaluation findings for each condition, namely noisy, enhanced, reverberant, dereverberation, and vocoded speech. Table \ref{table:detailedQ} and \ref{table:detailedI} present the comprehensive outcomes for speech quality and intelligibility, respectively. As observed, HASA-Net Large (WavLM-WS) consistently demonstrates superior performance compared to HASA-Net and HASA-Net Large (LS) across all conditions, as indicated by lower mean squared error (MSE) and higher correlation values. Remarkably, all three models exhibit similar trends in their performance. Notably, all three models exhibit similar patterns in their performance. Specifically, in terms of quality prediction, all models achieve higher correlations and lower mean squared error (MSE) values for enhanced and vocoded speech conditions compared to noisy, reverberation, and dereverberation conditions. Meanwhile, for intelligibility prediction, all models demonstrate better performance on reverberation and dereverberation speech conditions compared to noisy, enhanced, and vocoded conditions. 

\subsubsection{Evaluation of different fine-tuning}
We investigated the effects of different fine-tuning methods.
Based on its superior performance in quality and intelligibility predictions, we chose to use HASA-Net Large (WavLM-WS) for the subsequent fine-tuning experiments. For the fine-tuning process, we evaluate and contrast three distinct approaches: partial fine-tuning (PF), entire fine-tuning (EF), and a two-stage fine-tuning (2-stage FT) method as described in \cite{chen2022does}.
In the PF method, the convolutional feature extractor of the WavLM Large model is kept frozen, while only the transformers part is fine-tuned. On the other hand, in the EF approach, both the convolutional feature extractor and transformer layers are fine-tuned simultaneously during training. Regarding the 2-stage FT method, in the first stage, we keep the pretrained SSL model fixed and optimize the parameters of the remaining modules of HASA-Net Large. In the second stage, we fine-tune the entire HASA-Net Large, including the SSL model. Both in the PF and EF methods, we utilize the Adam optimizer with a learning rate of $10^{-5}$ for the SSL model, and a learning rate of $10^{-4}$ for the remaining modules of HASA-Net Large. In the 2-stage FT approach, we apply the Adam optimizer with a learning rate of $10^{-4}$ in the first stage, and a lower learning rate of $10^{-5}$ in the second stage. To the best of our knowledge, this is the first attempt to investigate the performance of apply various fine-tuning approaches in the context of a speech assessment model for individuals with hearing impairments.
\par The results of the fine-tuning methods on HASA-Net Large (WavLM-WS) are presented in Table \ref{table:finetuning}. The findings indicate that for quality prediction, the fine-tuning approaches consistently outperform the pretrained version. However, for intelligibility prediction, the PF approach exhibits slightly lower performance compared to the pretrained version. Overall, the 2-stage FT method demonstrates superiority over PF and EF, as evidenced by the lowest MSE and highest correlations, indicating better performance in both quality and intelligibility prediction.

\subsubsection{Transferability}
The transferability of HASA-Net Large on the OOD data was investigated. The OOD dataset consisted of training and testing sets with 23,100 (4,620 $\times$ 5) and 6,300 (1,260 $\times$ 5) utterances, respectively. A validation set of 460 $\times$ 5 utterances was randomly sampled from the training set. Every utterance was paired with an audiogram randomly selected from the corresponding training, validation, or testing set. This resulted in 4,160 $\times$ 5, 460 $\times$ 5, and 1,260 $\times$ 5 combinations for the training, validation, and testing sets, respectively.
\par We assessed the transferability of HASA-Net Large, including zero-shot, few-shot, and full-shot settings, where the number of available training examples varied. We utilized HASA-Net as our baseline approach once again. For the HASA-Net Large model, we opted for a 2-stage FT approach with WavLM Large-WS as the SSL representations, as it exhibited superior performance on the in-domain data. In the zero-shot setting, we directly assessed the models' performance on the OOD data. In the few-shot setting, we fine-tuned the models by using a limited amount of data, ranging from 100 to 12,800 examples. The quantity of data was increased by double for each successive amount. In the full-shot setting, we trained the models using the complete OOD dataset.
\par Table \ref{table:trans} shows that the zero-shot setting had the lowest performance compared to the few-shot and full-shot settings, which were less challenging. However, the HASA-Net Large model achieved moderate results for quality and intelligibility, despite the difficulty of the zero-shot setting. In the few-shot setting, both models demonstrated improvement with an increase in the amount of training data. Since the full-shot setting has complete access to the training data, it was anticipated that it would yield superior performance compared to the other two settings. It is noteworthy that both the HASA-Net and HASA-Net Large models revealed a greater improvement in quality estimation than in intelligibility estimation. Moreover, the HASA-Net Large consistently outperformed the HASA-Net model in different amounts of training data, highlighting its strong transferability and practical applicability in real-world scenarios. We also observe that with just 100 OOD training data, the HASA-Net Large model demonstrated a significant improvement in LCC for quality estimation, increasing from 0.765 to 0.951, and a smaller improvement in LCC for intelligibility estimation, increasing from 0.615 to 0.652. We also found that 3,200 utterances was adequate for achieving accurate quality and intelligibility predictions, as the performance of HASA-Net Large began to saturate at this point. Overall, the HASA-Net Large's impressive capacity to achieve excellent transferability with limited OOD training data makes it a highly valuable tool for practical applications.

\subsection{Characteristics of HASQI and HASPI}
\begin{figure}[t!]
  \centering
  \centerline{\includegraphics[width=8cm]{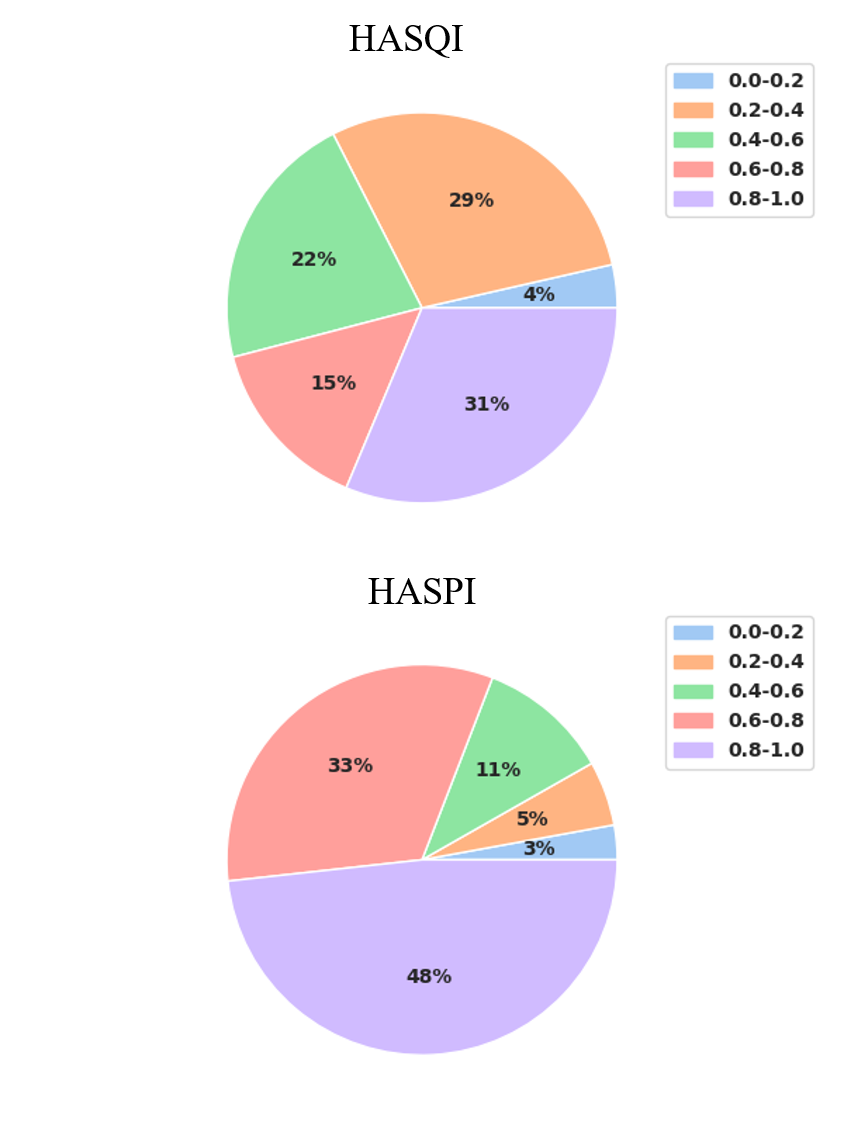}}
\caption{The percentage distribution of scores for HASQI and HASPI on the in-domain dataset, categorized into five intervals with a unit of 0.2 each.}
\label{fig:piechart}
\end{figure}

Our experimental results indicate that evaluating intelligibility is more challenging than evaluating quality, as demonstrated on both the in-domain VCTK dataset and the OOD TIMIT dataset. We observed that the differences in the sensitivity of the HASQI and HASPI metrics may be the primary cause of the inferior performance of intelligibility prediction. Previous studies such as \cite{kates2018using} also pointed out that HASQI is more sensitive to noise than HASPI. According to the findings in \cite{kates2018using}, HASPI approaches a value close to 1 at an SNR of 10 dB, while HASQI only reaches around 0.3 at the same SNR level. Even at a higher SNR of 40 dB, HASQI reaches a maximum of only 0.9. Notably, HASPI shows a transition from 0 to 1 between -10 dB and 10 dB, indicating that for most SNR levels above 10 dB, HASPI is expected to approach its maximum value. Our findings are in line with the observations made in our training data. Since the lowest SNR level in our datasets is 2.5 dB, this implies that the majority of HASPI scores are concentrated within the higher range of values. Fig. \ref{fig:piechart} displays the percentage distribution of scores for HASQI and HASPI on the in-domain dataset, with the scores divided into five intervals, each with a unit of 0.2.
Given the same waveform and audiograms, the distribution of HASQI scores appears to be fairly uniform across most values, with fewer scores falling within the range of 0 to 0.2. On the other hand, the distribution of HASPI scores is heavily skewed, with the majority of scores concentrated between 0.6 and 1.0. This suggests that there is limited data with low scores. The imbalanced distribution of HASPI scores in the training set presents challenges during the training process. Despite slightly lower prediction performance in terms of intelligibility compared to quality, HASA-Net Large shows promising results with high LCCs of 0.885 and \textbf{xxx} on the in-domain and OOD datasets, respectively.

\subsection{Comparison of SSL and Whisper models}

\begin{table*}[t]
    \caption{Comparison of Quality and Intelligibility Prediction Results between HASA-Net Large with Pre-trained WavLM and Whisper Model.}
    \centering
    \begin{tabular}{lcccccc}
    \toprule[0.4mm]    
    &\multicolumn{3}{c}{Quality}&\multicolumn{3}{c}{Intelligibility}\\
    \cline{2-4}\cline{5-7}
    Model & MSE$\downarrow$ & LCC$\uparrow$ & SRCC$\uparrow$ & MSE$\downarrow$ & LCC$\uparrow$ & SRCC$\uparrow$\\
    \hline  
    HASA-Net Large  & \multirow{2}{*}{\textbf{0.005}} & \multirow{2}{*}{\textbf{0.955}} & \multirow{2}{*}{\textbf{0.969}} & \multirow{2}{*}{\textbf{0.019}} & \multirow{2}{*}{\textbf{0.808}} & \multirow{2}{*}{\textbf{0.829}} \\
    (WavLM-WS) \\
    \hdashline
    HASA-Net Large  & \multirow{2}{*}{0.008} & \multirow{2}{*}{0.951} & \multirow{2}{*}{0.942} & \multirow{2}{*}{0.026} & \multirow{2}{*}{0.740} & \multirow{2}{*}{0.732} \\
    (Whisper) \\
    \bottomrule[0.4mm]
    \label{table:w/whisper}
    \end{tabular}%
\end{table*}

While our study primarily centered on employing SSL models for speech assessment, an impressive weakly-supervised model named Whisper \cite{radford2022robust} has emerged. Whisper is trained using a total of 680,000 hours of labeled audio and has shown outstanding performance on various ASR tasks. Therefore, we are also intrigued by the possibility of integrating Whisper's representations into our HASA-Net Large model. To the best of our
knowledge, this is the first work to integrate Whisper into speech assessment.

\par Whisper first computes the log-Mel spectrogram, which is then passed through two convolutional layers, followed by transformer layers. The raw waveform is fed into Whisper, and we extract the representations obtained after the transformer layers as the input for the HASA-Net Large. From the results in Table \ref{table:w/whisper}, despite the superior performance achieved by using wavlm representations, integrating Whisper representations results in a slightly lower outcome, specifically with a decrease of 0.004 and 0.068 in correlation values for quality and intelligibility, respectively. Furthermore, when comparing it to HASA-Net, the incorporation of Whisper representations into HASA-Net Large yields improved results. This is evident from Table \ref{table:pretrain}, which shows that HASA-Net achieves lower correlation values of 0.947 and 0.737 for quality and intelligibility, respectively, while incorporating Whisper leads to enhanced performance. These results further confirm the effectiveness of incorporating Whisper representations for speech assessment.

\section{Conclusion}
\label{sec:conclusion}
In this paper, we introduce HASA-Net Large, a multi-objective
non-intrusive hearing-aid speech assessment model, which builds upon previous work on HASA-Net. HASA-Net Large improves upon HASA-Net in several ways. First, it is a general model that takes into account both normal-hearing and hearing-impaired listeners. Secondly, it incorporates SSL pretraining and fine-tuning techniques, demonstrating the superiority of SSL in enhancing the model's performance. Thirdly, it evaluates the robustness of the model across five different speech conditions, including noisy, enhanced, reverberation, dereverberantion, and vocoded speech. Fourthly, the transferability of the model is validated using the OOD dataset in zero-shot, few-shot, and full dataset scenarios. The experimental results reveal that incorporating SSL model yields better performance compared to the previous HASA-Net, which used spectrogram as input features. We also explored various fine-tuning approaches for quality and intelligibility prediction tasks. Moreover, HASA-Net Large demonstrates improved transferability under various training data sizes, indicating its applicability in real-world scenarios. We also identify the difficulty posed by the imbalance in the distribution of intelligibility scores, which is an ongoing effort to address in order to improve prediction accuracy. These findings validate that the proposed HASA-Net Large has the potential to serve as a universal model for practical applications.

\bibliographystyle{IEEEtran}
\bibliography{mybib}

\end{document}